\newcommand{\singlespace}{
        \renewcommand{\baselinestretch}{1}
        \large\normalsize}

\def\dirac#1{\hbox to 0pt{\slash}{#1}}

\documentstyle[aps,prc,preprint,floats,epsfig]{revtex}
\def\dirac#1{\hbox to 0pt{\slash}{#1}}

\begin{document}
\singlespace

\title{Real Compton Scattering at High $p_\perp$}

\author{Alan M. Nathan\\
University of Illinois at Urbana-Champaign\\
(for the E97-108 collaboration)\\
Contribution to Jefferson Lab Workshop on \\
Physics \& Instrumentation with 6-12 GeV Beams}

\maketitle

\narrowtext
\singlespace


\vspace{0.25in}
Compton scattering ($\gamma + p \rightarrow \gamma + p$) in the hard scattering
limit
is a potentially powerful probe of the short-distance structure of the nucleon.
It is a natural complement to high $Q^2$ exclusive reactions, such as elastic form factors and
deeply virtual Compton scattering (DVCS),
where the common feature 
is a hard energy scale, leading to a
factorization of the transition amplitude into a part involving the overlap of soft
(nonperturbative) wave functions and a perturbative hard scattering amplitude 
(HSA).
For Real Compton Scattering (RCS), the hard scale is achieved when both $s$
and the transverse momentum transfer $p_\perp$ are large, and under these
conditions RCS is sometimes referred to as ``Wide-Angle Compton Scattering.''  
There are two
types of questions that are interesting to investigate:
\begin{enumerate}
\item What is the appropriate mechanism for the HSA?
\item What can RCS teach us about the proton wave function?
\end{enumerate}
In this report, we discuss the physics motivation for such measurements in light of
these two questions.  We then discuss some of the experimental considerations for a program
at an upgraded JLab facility.  Measurements up to 6 GeV are already in the planning stage (E97-108)
[1].

Different HSA mechanisms can be distinguished by the manner in which 
$p_\perp$ is shared among the constituents.  Asymptotically this sharing
is expected to occur in the HSA via hard gluon exchange (Fig. 1a), 
\begin{figure}[bt]
\begin{center}
\epsfig{angle=0,width=6in,file=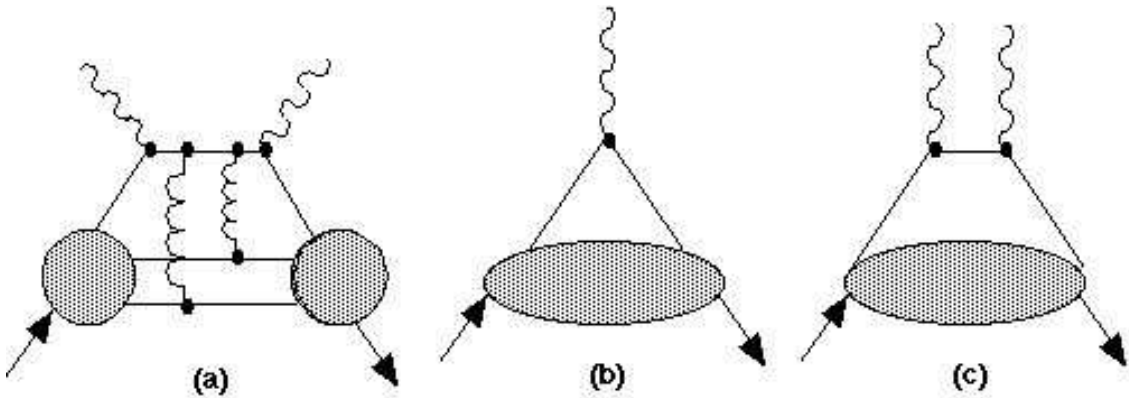}
\caption{Two-gluon exchange diagram for RCS (a) and handbag diagram for
form factors (b) and RCS (c).}
\end{center}
\end{figure}
leading to the quark counting rule and scaling~[2],
\begin{eqnarray}
{d\sigma / dt} &=& \frac{f(\theta_{cm})}{s^n},
\label{eq:QuarkCounting}
\end{eqnarray}
where n=6 for RCS.  With modest precision, existing data from Cornell [3] 
approximately support scaling with n=6
(see Fig.~2a).  
\begin{figure}[hbt]
\begin{center}
\epsfig{angle=0,width=6in,file=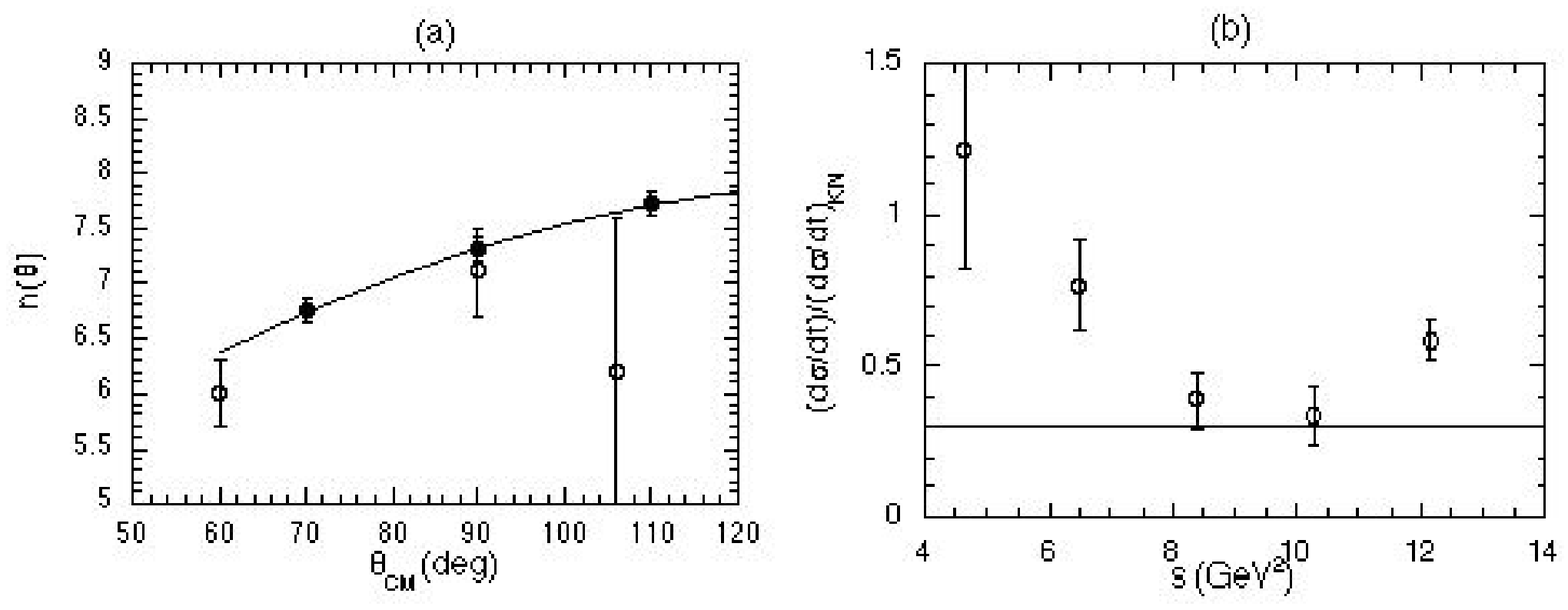}
\vspace{0.25in}
\caption{(a) Scaling of RCS cross section at fixed $\theta_{cm}$.  The
open points are the Cornell data and the closed points are the projected
results from E97-108.  The curve is the prediction of Radyushkin, assuming
dominance of the handbag diagram.  (b) Ratio of the Cornell cross section to KN at
fixed $t=-2.45 \, GeV^2$.  The horozontal line is the prediction of Radyushkin based on
a model for the form factor and shows the level of agreement with the existing data.}
\end{center}
\end{figure}
Nevertheless, Radyushkin [4]
and others argue that this mechanism, when combined with realistic wave functions, badly
underpredicts the RCS cross section.  Instead he suggests that 
the dominant mechanism at experimentally accessible energies for both form factors and RCS
is the handbag diagram (Fig. 1b,c),
whereby the large $p_\perp$
is absorbed on a single quark and shared 
by the overlap of high momentum components in the soft wave function.  The latter
are described in terms of the same nonforward parton densities (NFPD) that appear in
deep inelastic scattering and DVCS, thereby offering the possibility of a unified description of
all these processes.
The dominance of the handbag diagram leads
to the approximate factorization of the RCS cross section into the
product of the Klein-Nishina (KN) cross section and a form factor-like object:
\begin{eqnarray}
{d\sigma\over dt} \approx 
\left ({d\sigma\over dt}\right )_{\rm{KN}} | F_{\gamma\gamma}(t) |^2
& \phantom{space}& F_{\gamma\gamma}(t)=
 \sum_f e_f^2 \int_0^1 F_f(x,t)\frac{dx}{x} ,
\label{eq:Feynman}
\end{eqnarray}
where the function  $F_f(x,t)$ is a particular projection
of the NFPD for quark flavor $f$.   This factorization leads to interesting consequences that can
be tested experimentally.  First, the 
the scaling parameter n in Eq.~1 is angle dependent, as shown in Fig.~2a.
Moreover the ratio $\sigma/\sigma_{KN}$ is $s$-independent
at fixed $t$, in approximate accord with the data (Fig.~2b).  Further, the ratio
yields F$_{\gamma\gamma}(t)$, which is a new form factor about which we know very
little experimentally and which
contains the interesting physics about the soft wave function. 
It is similar to but different from
the Pauli form factor $F_1(t)$ measured in elastic electron scattering:
$$F_1(t)=\sum_f e_f \int_0^1 F_f(x,t)dx. $$  
For example,  $F_{\gamma\gamma}(t)$ is enhanced relative to
$F_1(t)$ due to the 1/x weighting in the integral.  Moreover, the weighting by the
square of the quark charge means that $F_{\gamma\gamma}(t)$ is sensitive to
the flavor structure of the proton in a different way from $F_1(t)$,
thereby providing another possible tool (along with parity-violating electron
scattering) for decomposing the flavor structue.  In particular, RCS is dominated by
the $u$-quark distribution.

We now give some brief remarks about measurement with circularly polarized incident
photons and a polarized proton target.  The
same physics is probed with
an unpolarized target and measurement of the recoil polarization.
In the absence
of theoretical work in this area , we use the polarization-dependent KN cross
section for guidance, for which the beam asymmetry
\begin{equation}
S_z\,\equiv\,\frac{\sigma(j=1/2)-\sigma(j=3/2)}{\sigma(j=1/2)+\sigma(j=3/2)}
\end{equation}
is large (e.g., $\sim$0.7 at 
E=6 GeV and $\theta_{cm}$=90$^\circ$).  This suggests that polarized photons
selectively scatter from quarks polarized in the opposite direction from
the photon.  Thus we have the exciting possibility that polarized RCS
is sensitive to the spin-flavor structure of the generalized form factor,
especially for the valence $u$ quarks.  Because of luminosity limitations with
a polarized target or the small figure-of-merit with focal-plane polarimeters,
these double-polarization experiments
are only feasible when the scattering cross sections are large, 
i.e. for $s<8$ and $-t<4$ GeV$^2$.

Experimentally, one would like to test our understanding of the HSA
mechanism by measuring RCS cross sections with good precision over
a broad range of $s$ and $t$ in order to check the scaling behavior at both
fixed $\theta_{cm}$ and fixed $t$.  Moreover, one would like
to measure $F_{\gamma\gamma}(t)$ over a
similar range as $F_1(t)$.  
In Fig.~3a we show the kinematics appropriate to a 12 GeV upgrade of JLab,
as well as limitations imposed by existing magnetic spectrometers and
by the hard scattering condition $p_\perp\geq$ 1 GeV.  The dots show the
proposed kinematics for E97-108, for which the maximum incident energy
is 6 GeV.  We note that
the only relevant data [3] extend up to
$s$=12, $-t$=6 GeV$^2$ but are of poor quality
above $s$=8. 
 
\begin{figure}[hbt]
\begin{center}
\epsfig{angle=0,width=6in,file=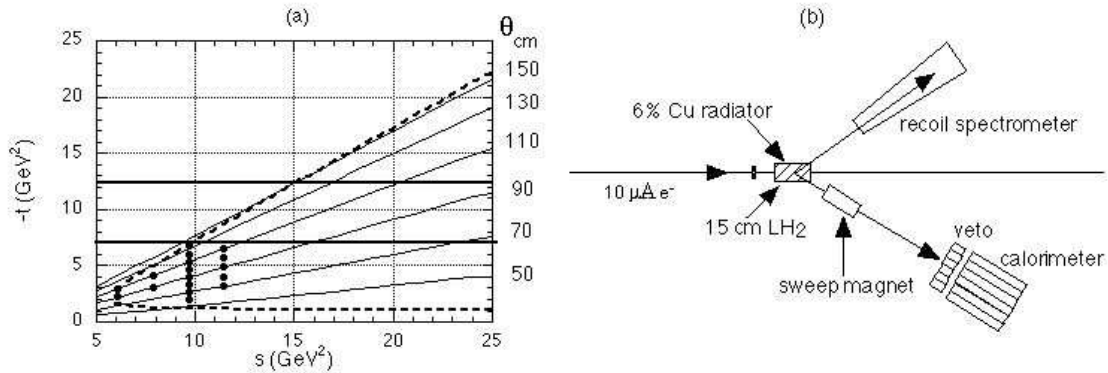}
\vspace{0.25in}
\caption{(a) Kinematics of RCS measurements up to 12 GeV.  The solid diagonal lines are contours at
fixed $\theta_{cm}$; the dashed diagonal lines indicate the upper and lower angular limits
for $p_\perp\geq$1 GeV; the bold horizontal lines indicate the upper limit of the existing Hall A
(4.5 GeV/c) and Hall C (7.5 GeV/c) spectrometers for the recoil proton.  The dots are the
proposed kinematics of E97-108.  (b) Plan view of a possible RCS setup.}
\end{center}
\end{figure}

An overview of the experimental apparatus for RCS is shown in Fig.~3b.  
The small cross sections require high luminositiy, which precludes the
use of tagged photons for precision measurements.
Instead a high intensity ($\geq$ 10 $\mu$A) beam
of electrons impinges on a 6\% copper
radiator, and the {\it mixed} electon-photon beam is 
incident on a 15-cm LH$_2$ scattering target.
For incident photons
near the bremsstrahlung endpoint,
the recoil proton and scattered photon
are detected
with high angular precision in a magnetic spectrometer and photon
spectrometer, respectively.
An essential feature is to use the kinematic correlation
between the scattered photon and recoil proton in the $p(\gamma,\gamma^\prime p)$
reaction to reduce
the copious background of photons from the decay of $\pi^0$'s
from the $p(\gamma,\pi^0 p)$ reaction, thereby placing stringent
demands on the
angular resolution of each spectrometer.  

The photon spectrometer for E97-108 is currently being designed, but the important
compontents have already been identified.  First one needs
a large-area segmented calorimeter with modest energy resolution and excellent
position resolution.
An array of 625 Pb-Glass blocks will be used,
each with dimensions 4x4x40 cm$^3$ and with an expected position resolution of order 5 mm. 
Next, electrons from $ep$ scattering are deflected by a magnet and 
identified in a scintillator hodoscope, effectively removing this source
of background.
The magnet also
serves to sweep away low-energy electrons, thereby reducing the total energy flux on the
detector.  Finally
a MWPC in front of the calorimeter will be used in a separate {\it in situ}
calibration experiment to
measure the position resolution of the calorimeter using $ep$ electrons.  The entire spectrometer
will be mounted on a mechanical assembly that will allow changes in both
scattering angle and radial distance,
the latter needed to match the photon acceptance to that of the proton at different
kinematic settings.

In summary, there is much exciting physics to be learned from 
RCS at high $p_\perp$.
The general technique described here for E97-108
should work with only minor changes
at an upgraded JLab facility,
with event rates ranging from a few thousand/hour at 3 GeV to a few hundred/hour at 5 GeV
to about 10/hour at 12 GeV.  One can therefore look forward to a vigorous
program at JLab extending well into the next decade.

\begin{center}
{\Large References}\
\end{center}
\noindent
1.  C. Hyde-Wright, A. M. Nathan, and B. Wojtsekhowski (co-spokespersons),
JLab Proposal E97-108 (1997).\\
2.  S. J. Brodsky and G. Farrar, Phys. Rev. Lett. {\bf 31}, 1953 (1973).\\
3.  M. A. Shupe {\it et al.}, Phys. Rev. D {\bf 19}, 1929 (1979).\\
4.  A. Radyushkin, hep-ph/9803316 (1998).

\end{document}